\documentclass[letter]{jpsj3-cond-mat}
\usepackage{graphicx}
\usepackage{txfonts}
\usepackage{color}

\title{
Giant Magnetoresistance Effect in the Metal-Insulator Transition of Pyrochlore Oxide Nd$_2$Ir$_2$O$_7$
}

\author{
Kazuyuki {\sc Matsuhira}$^{1}$\thanks{E-mail address: matuhira@elcs.kyutech.ac.jp}, Masashi {\sc Tokunaga}$^{2}$, Makoto {\sc Wakeshima}$^{3}$, Yukio {\sc Hinatsu}$^{3}$, and Seishi {\sc Takagi}$^{1}$
}

\inst{
\address{
$^1$Graduate School of Engineering, Kyushu Institute of Technology, Kitakyushu 804-8550, Japan\\
}
\address{
$^2$ Institute for Solid State Physics, University of Tokyo, Kashiwa 277-8581, Japan\\
}
\address{
$^3$Division of Chemistry, Graduate School of Science, Hokkaido University, Sapporo 060-0810, Japan\\
}}

\abst{
We investigated the magnetoresistance (MR) effect of the pyrochlore oxide Nd$_2$Ir$_2$O$_7$, which shows a metal-insulator transition at $T_{\rm MI}$ = 33 K.
A small positive MR effect was observed in the metallic state above $T_{\rm MI}$, while a 
large negative MR effect was observed in the insulating state below $T_{\rm MI}$.
MR effects ($[\rho(B = 0) - \rho(B)]/\rho(B)$) exceeding 3000\% were found at 1 K at a field of 9 T.
As a result, we confirmed the crossover from the insulating state to a state with a small or partial band gap in a field up to 56 T.
Furthermore, from the MR effect in Eu$_2$Ir$_2$O$_7$ ($T_{\rm MI}$ = 120 K) and Gd$_2$Ir$_2$O$_7$ ($T_{\rm MI}$ = 127 K), we revealed that the large negative MR effect of the pyrochlore iridate {\it Ln}$_2$Ir$_2$O$_7$ depends on the magnetism of the lanthanide {\it Ln}$^{3+}$ ion.
The {\it d-f} interaction plays a significant role in the large negative MR effect in the insulating state.
}

\kword{
magnetoresistance, pyrochlore oxides, Nd$_2$Ir$_2$O$_7$, metal-insulator transition, Eu$_2$Ir$_2$O$_7$, Gd$_2$Ir$_2$O$_7$
}

\newpage

\begin{document}
\maketitle

The discovery of giant magnetoresistance (GMR) in thin-film structures composed of alternating ferromagnetic and nonmagnetic layers has opened up a new field of electronics called spintronics.~\cite{Novel}
The GMR effect is observed as a large change in resistance depending on the parallel or antiparallel alignment of  ferromagnetic layer magnetizations.
This effect has been applied in disk read heads of modern hard disk drives and magnetic sensors.
However, GMR has also been discovered  in strongly correlated electron systems, mainly in  manganites.~\cite{Tokura, Tl2Mn2O7}
This GMR is caused by the field-induced transition from nonmetal to metal as a bulk property.
This effect is qualitatively explained by carrier scattering due to the thermally fluctuating spin configuration or the spin-disorder scattering process.

A recent study has revealed a metal-insulator transition (MIT) in 4{\it d} and 5{\it d} transition-metal pyrochlore oxides.~\cite{CdOs1, CdOs2, LnIrKM1, LnIrKM2, HgRu}
Pyrochlore oxides are composed of a network of corner-shared tetrahedra, whose vertices are occupied by spins; this is called the pyrochlore lattice.~\cite{Pyrochlore}
As these spins may give rise to strong geometrical frustration, unusual spin fluctuations are underlying in pyrochlore oxides.~\cite{AndersonSP, TbMo, YMo, SpinIce, KI}
Therefore, 4{\it d} and 5{\it d} transition-metal pyrochlore oxides can indicate novel electronic properties such as GMR in a magnetic field.

In our recent study, we successfully synthesized purified polycrystalline samples of the pyrochlore iridate {\it Ln}$_2$Ir$_2$O$_7$.~\cite{LnIrKM1, LnIrKM2}
We revealed that {\it Ln}$_2$Ir$_2$O$_7$ for {\it Ln} = Nd, Sm, Eu, Gd, Tb, Dy, and Ho exhibits MITs at 33, 117, 120, 127, 132, 134, and 141 K, respectively.~\cite{LnIrKM2}
In these MITs, thermal hysteresis and a discontinuous change in the physical properties of pyrochlore iridates was not observed at the MIT temperature $T_{\rm MI}$.
From this result, we concluded that the MITs were second-order transitions.
The MIT comes from $5d$ electrons.
In a recent theoretical study, on this insulating state of {\it Ln}$_2$Ir$_2$O$_7$, the possibility of realizing a topological insulator is discussed.~\cite{Ir-Balents}

In this letter, we report on the magnetoresistance (MR) of Nd$_2$Ir$_2$O$_7$, which shows an MIT at 33 K.
This MIT involves the magnetic ordering of Ir moments.
The all-in/all-out state in the insulating state has been revealed by neutron scattering experiments.~\cite{NdIrTomiyasu}
Furthermore, a long-range magnetic structure of the Nd moment induced by {\it d-f} interaction develops below 15 K.
The magnetic structure of the Nd moment is also an all-in/all-out state.
This antiferromagnetic structure is destroyed by applying a magnetic field.
Therefore, we may expect the transition from the insulating state with antiferromagnetic ordering to a polarized metallic state.
Furthermore, as the origin of the MIT in {\it Ln}$_2$Ir$_2$O$_7$ is still not clear, an  investigation of the magnetic field response may provide useful information on the origin of the MIT.

Polycrystalline samples of {\it Ln}$_2$Ir$_2$O$_7$ ({\it Ln} = Nd, Eu, Gd) and (Nd$_{0.70}$Pr$_{0.30}$)$_2$Ir$_2$O$_7$ were synthesized by a standard solid-state reaction method according to previously reported preparation conditions.~\cite{LnIrKM2}
The resistivity measurement was performed by a four-probe method.
The DC resistivities up to 9 T were measured under a static magnetic field (Quantum Design PPMS). 
The DC and AC resistivities up to 56 T were measured using a pulsed magnetic field at the Institute for Solid State Physics, the University of Tokyo.
In all MR measurements, the magnetic field was applied parallel to the current direction.

Figure 1(a) shows the MR of Nd$_2$Ir$_2$O$_7$ at 1 K under a static magnetic field.
Before the initial up-sweep of the magnetic field, the sample was cooled under a zero field.
In the initial up-sweep of the magnetic field, the resistivity decreases at low fields and exhibits a hump at around 4.5 T; a positive MR is observed in the magnetic field range of 2-4.5 T.
In the down-sweep of the magnetic field down to 0 T, the resistivity increases monotonically.
Therefore, a large hysteresis is observed; in a zero field, the resistivity after the first sweep becomes 17.9 times larger.
Consequently, an MR effect ($[\rho(B = 0) - \rho(B)]/\rho(B)$) exceeding 3000\% is found at 1 K at a magnetic field of 9 T.
In the second up-sweep, a small hysteresis is observed without a hump.
The large negative MR effect observed in the ordered state may be qualitatively explained by the suppression of carrier scattering due to the fluctuating spin configuration with increasing field.

Figure 1(b) shows the MR and the magnetization process of Nd$_2$Ir$_2$O$_7$ at 2 K under a static magnetic field.
The MR at 2 K indicates a qualitatively similar feature to that observed at 1 K; a large hysteresis  corresponding to curves 1 and 2 in Fig. 1(a) is observed.
The magnetization process at 2 K shows a small hysteresis at around 3 T.
In addition, a tiny residual magnetization of 10$^{-3}$ $\mu_{\rm B}$/f.u. is observed; although the residual magnetization appears as a tiny ferromagnetic moment below $T_{\rm MI}$, the origin is not still clear.
Therefore, we can speculate that the large hysteresis at a zero field is caused by the tiny ferromagnetic moment.
These results suggest that the large hysteresis in the MR is related to the small hysteresis in the magnetization process.

We will now discuss the relationships between the MR and magnetization from the results shown in Fig. 1(b).
The magnetization $M$ of Nd$_2$Ir$_2$O$_7$ reaches 2.3 $\mu_{\rm B}$/f.u. at 9 T.
The saturated magnetization $M_{\rm s}$ is estimated to be 2.5 $\mu_{\rm B}$/f.u. by Brillouin curve fitting.
As the magnetic moment is mainly caused by Nd moments, the saturated magnetization per Nd ion may be described by 1.25 $\mu_{\rm B}$/Nd.
Because the Nd moment has local $\langle 111\rangle$ Ising anisotropy, the saturated magnetization is suppressed to half of the single-ion moment.
From the analysis of inelastic neutron scattering, the value of the Nd moment of the ground state doublet is 2.37 $\mu_B$.
Therefore, this estimated Nd moment is in agreement with the result of inelastic neutron scattering.
First, in the down-sweep, we found that ${\rm ln} \rho \propto (M/M_{\rm s})^2$ at $(M/M_{\rm s})^2 \leq 0.2$, where $M_{\rm s}=2.5$ $\mu_{\rm B}$/f.u.; this feature has been discussed for the charge-ordered state of Mn oxides (La, Sr)$_2$MnO$_4$.~\cite{Tokunaga}
Alternatively, we can obtain the change in resistivity ($\Delta \rho/\rho _0$) in the up- and down-sweeps as a function of $M^2$, where $\Delta \rho/\rho _0 = [\rho(B) - \rho(B = 0)]/\rho(B=0)$.
In the small-magnetization region ($(M/M_{\rm s})^2 \leq 0.04$), we found that both are proportional to $M^2$ with different coupling constants; the feature has been discussed for Mn perovskites (La, Sr)MnO$_3$.~\cite{Tokura, Furukawa, CC}
However, as is clear from the result shown in Fig. 1(b), it is hard to explain the discrepancy between the large hysteresis in the resistivity and the small hysteresis in the magnetization from the above-mentioned relationships.
Further study is needed to clarify the origin of this phenomenon.

\begin{figure}[htbp]
\begin{center}
\includegraphics[width=7.5cm]{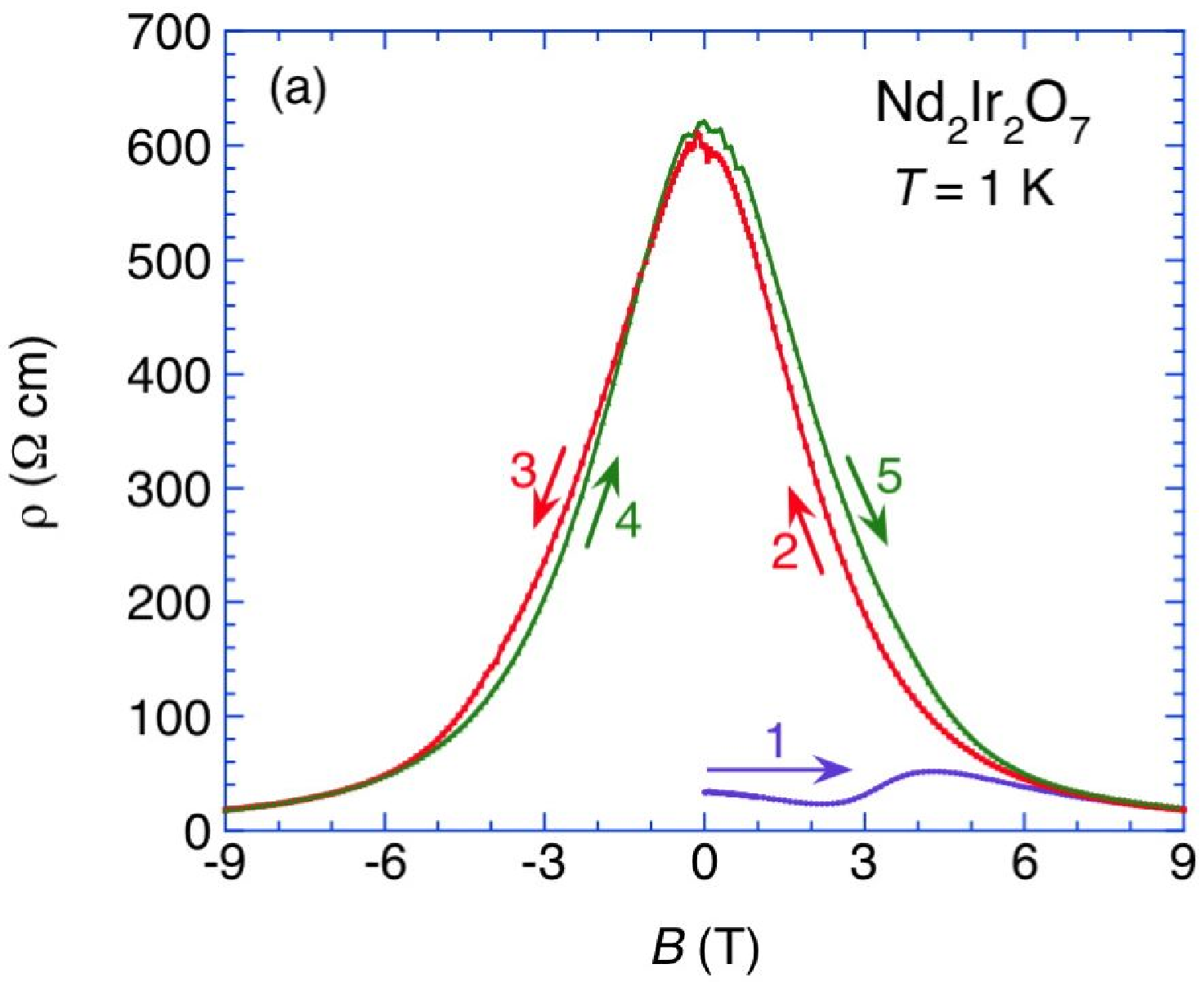}
\includegraphics[width=7.5cm]{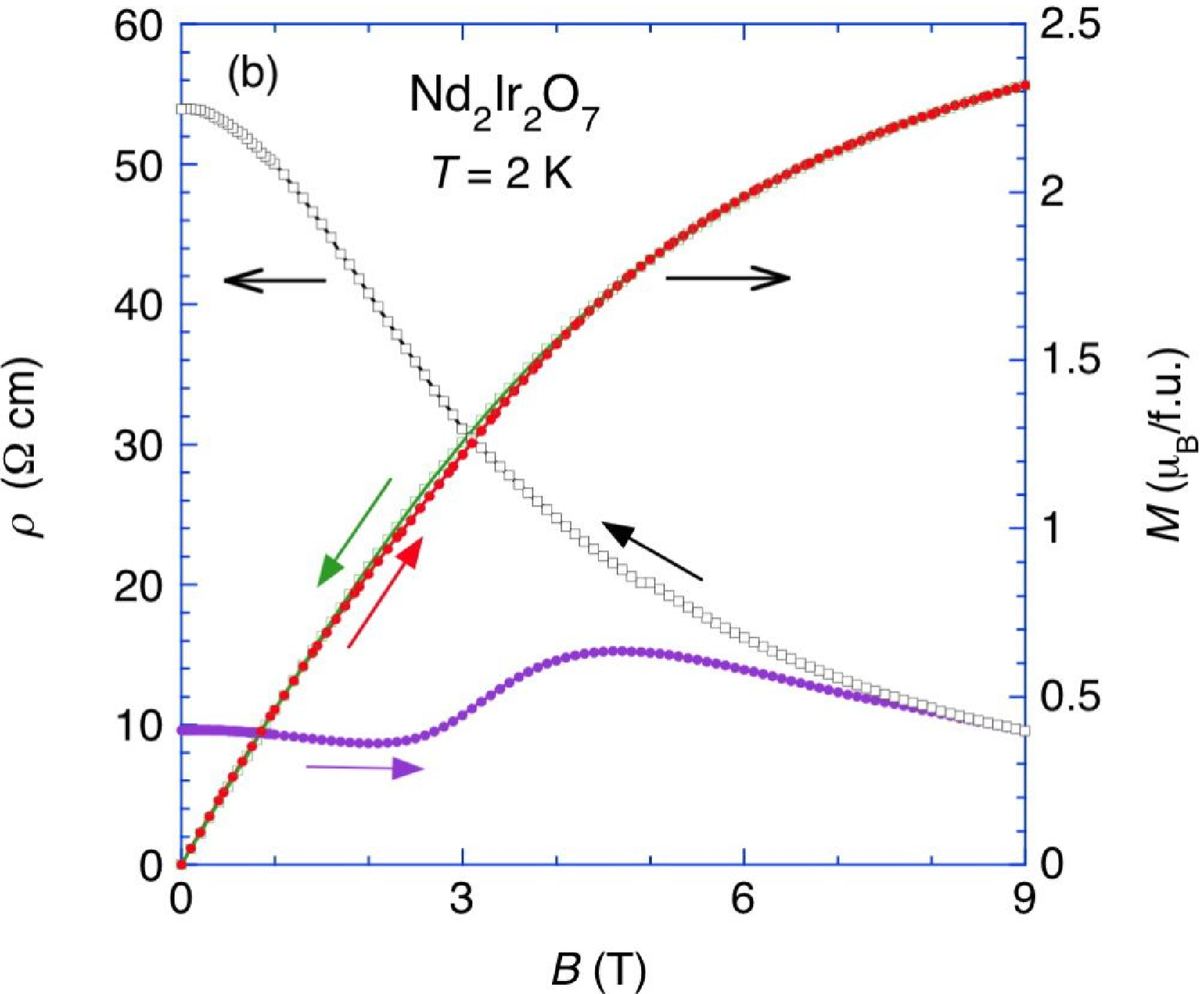}
\end{center}
\caption{
(a) (Color online) Magnetoresistance of Nd$_2$Ir$_2$O$_7$ at 1 K under a static magnetic field. The purple line shows the first up-sweep. The red line shows the down-sweep. The green  line shows the second up-sweep. (b) (Color online) Magnetoresistance and magnetization process of Nd$_2$Ir$_2$O$_7$ at 2 K under a static magnetic field.
}
\label{f1}
\end{figure}
 
Figure 2 shows the MR of Nd$_2$Ir$_2$O$_7$ under pulsed magnetic fields after an initial field sweep.
At 40 K (above $T_{\rm MI}$), a small positive MR effect was observed up to 56 T.
However, below $T_{\rm MI}$, a negative MR effect was observed up to 56 T, which can be expected from the results shown in Fig. 1.
At 4.2 K, the ratio of resistivity $\rho(B)/\rho(0 {\rm T})$ between 0 and 56 T reaches 0.007.
However, there is no sign of a phase transition to the metallic state.
The resistivity at 56 T increases with decreasing temperature.
For the data at 10, 15, and 25 K, a band gap of 25 K is estimated at 56 T assuming thermally activated conduction (see the inset in Fig. 2); with the same assumption, for the data just below $T_{\rm MI}$, a band gap of 405 K is obtained at a zero field.~\cite{LnIrKM2}
A state with a small or partial band gap is realized at a field of 56 T.

Next, we will discuss the large hysteresis in the medium temperature range (at 10, 15, and 25 K).
This is caused by a magnetocaloric effect due to Nd moments because the sample is set in a quasi-adiabatic condition; in the sweep at 4.2 K, because the sample is soaked in liquid $^4$He, it is set in an isothermal condition.
Assuming an adiabatic condition in the up-sweep, the increase in sample temperature at 10 K is estimated to be approximately 15 K.
The sample temperature unavoidably increases in the up-sweep.
The actual increase in sample temperature is expected to be 2-3 K.
On the other hand, the sample is cooled in the down-sweep.
Actually, after the field sweep, a relaxation to the thermal equilibrium state is observed as the result of a decrease in resistivity.
Therefore, this hysteresis in the medium temperature range is not intrinsic.

\begin{figure}[htbp]
\begin{center}
\includegraphics[width=7.5cm]{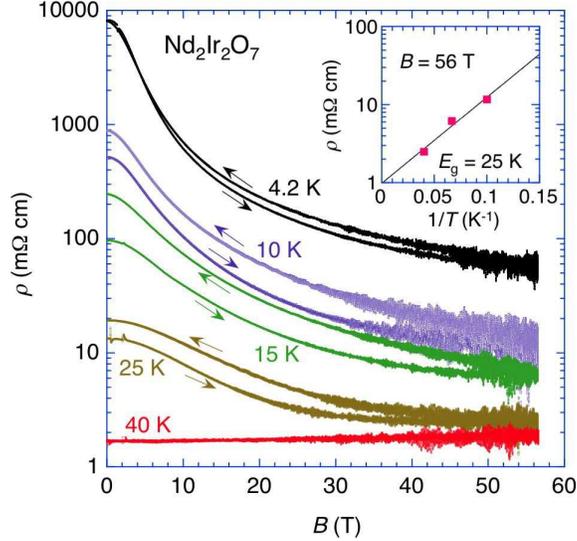}
\end{center}
\caption{
(Color online) Magnetoresistance of Nd$_2$Ir$_2$O$_7$ under pulsed magnetic fields after an initial field sweep. The inset shows the temperature dependence of resistivity at 56 T plotted against $1/T$.
}
\label{f2}
\end{figure}

To clarify the insulating state under a high magnetic field, we measured the MR of a system with a lower $T_{\rm MI}$.
Figure 3 shows the MR of (Nd$_{0.70}$Pr$_{0.30}$)$_2$Ir$_2$O$_7$ under pulsed magnetic fields after an initial field sweep.
The inset shows the temperature dependence of the resistivity of (Nd$_{0.70}$Pr$_{0.30}$)$_2$Ir$_2$O$_7$.
This compound exhibits an MIT at $T_{\rm MI}$ = 21 K.
At 25 K, a small positive MR effect is observed at fields of up to 56 T.
Below $T_{\rm MI}$, a negative MR effect is observed at fields of up to 56 T.
At a field of 56 T, the resistivity at 1.4 K  is 10 times larger than that at 25 K.
These features are consistent with the results for Nd$_2$Ir$_2$O$_7$.
There is no sign of a phase transition to the metallic state up to 56 T.

Next, we will discuss the Zeeman energy in the insulating state.
Assuming the Ir moment of $1 \mu_{\rm B}$, the Zeeman energy at a field of 56 T is estimated to be 37.5 K.
For Nd$_2$Ir$_2$O$_7$, the Zeeman energy is comparable to the energy gain of the ordered state ($T_{\rm MI}$ = 33 K).
Furthermore,  for (Nd$_{0.70}$Pr$_{0.30}$)$_2$Ir$_2$O$_7$, the Zeeman energy at a field of 56 T is greater than  the energy gain of the ordered state ($T_{\rm MI}$=21 K).
However, the present result shows that the insulating state persists up to 56 T.
There are two possible interpretations for the insulating state persisting in a high magnetic field.
One is the reduction in the value of the Ir moment.
We will try to roughly estimate the value of the Ir moment as shown below.
The resistivity at 4.2 K shows a field dependence of $H^{\alpha}$ above 30 T, where $\alpha = -1.28$.
The resistivity at 40 K shows the field dependence of $\rho_0 + \beta H^2$.
Their extrapolated curves cross at around 300 T; this is considered as the critical field.
By using this critical field, we can obtain a large reduction in the moment of 0.16 $\mu_{\rm B}/{\rm Ir}$ from the Zeeman energy for the critical field.
This strong reduction in the magnetic moment may be caused by a frustration effect. 
Another interpretation is that the band gap formation does not originate from magnetic ordering; this means that the MIT is a Mott transition with a magnetic ordering rather than a Slater transition.
In this case, the band gap may originate from on-site Coulomb repulsion and spin-orbit coupling.
Because the energy scale (of order eV) is large, it is difficult to destroy the band gap completely by a magnetic field.
To clarify this point, further study is needed to estimate the ordered Ir moment.

\begin{figure}[htbp]
\begin{center}
\includegraphics[width=7.5cm]{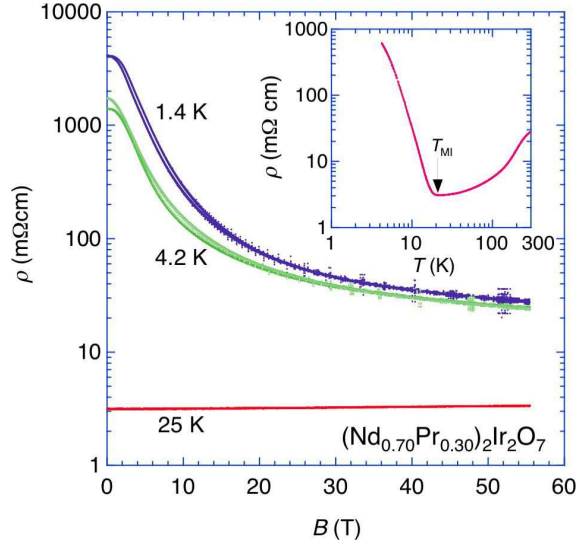}
\end{center}
\caption{
(Color online) Magnetoresistance of (Nd$_{0.70}$Pr$_{0.30}$)$_2$Ir$_2$O$_7$ under pulsed magnetic fields. The  inset shows the temperature dependence of resistivity.
}
\label{f3}
\end{figure}

Figure 4 shows the MR of Eu$_2$Ir$_2$O$_7$ and Gd$_2$Ir$_2$O$_7$ under pulsed magnetic fields after an initial field sweep.
The $T_{\rm MI}$ for Eu$_2$Ir$_2$O$_7$ and Gd$_2$Ir$_2$O$_7$ are 120 and 134 K, respectively.~\cite{LnIrKM2}
The nonmagnetic Eu$^{3+}$ ion shows Van Vleck susceptibility.
On the other hand, the Gd$^{3+}$ ion has an isotropic magnetic moment.
Eu$_2$Ir$_2$O$_7$ shows a small positive MR effect above and below $T_{\rm MI}$.
However, the MR effect of Gd$_2$Ir$_2$O$_7$ is small positive and negative above and below $T_{\rm MI}$, respectively.
These results suggest that the negative MR effect of {\it Ln}$_2$Ir$_2$O$_7$ below $T_{\rm MI}$ is related to the magnetic moment at the {\it Ln} site.
An effect caused by {\it d-f} interaction is suggested.

Now, we discuss the MR effect of Nd$_2$Ir$_2$O$_7$.
The magnetic structure of the Ir moment in the insulating state is all-in/all-out.
Below $T_{\rm MI}$, because the conductivity is caused by an excitation from the all-in/all-out state, the mechanism of conductivity must be variable range hopping (VRH).~\cite{VRH}
Actually, Nd$_2$Ir$_2$O$_7$ obeys the VRH conduction formula below 15 K ($\sim T_{\rm MI}/2$):~\cite{LnIrKM2} polycrystalline Eu$_2$Ir$_2$O$_7$ also shows VRH conduction below 60 K ($\sim T_{\rm MI}/2$).~\cite{LnIrKM2,EuIrIshikawa}
The strength of the {\it d-f} interaction is roughly 10 K in this system.
Below $T_{\rm MI}$, the {\it d-f} interaction leads to a finite internal field at the Nd sites.~\cite{LnIrKM2,NdIrTomiyasu}
The VRH conduction may lead to a fluctuation of the internal field at the Nd sites through the {\it d-f} interaction.
As the Nd moment fluctuates, disorder is created in the magnetic structure of the Nd moment.
In a magnetic field, because the Ir moments are small, the energy gain due to the Zeeman effect is small.
However, because the Nd moments are larger, the energy gain due to the Zeeman effect is larger.
As the applied magnetic field increases, the magnetic structure of the Nd moments changes from the all-in/all-out state to the 2-in/2-out or 1-in/3-out (3-in/1-out) state that the magnetic moment is finite.
Then, the Nd moments begin to align in the magnetic field direction against the internal field owing to the {\it d-f} interaction.
Therefore, the disorder in the magnetic structure of the Nd moments is suppressed by applying a magnetic field.
Consequently, because the fluctuation of the internal field at the Ir sites due to the {\it d-f} interaction is suppressed in a stronger magnetic field, the MR effect may be negative.

\begin{figure}[htbp]
\begin{center}
\includegraphics[width=7.5cm]{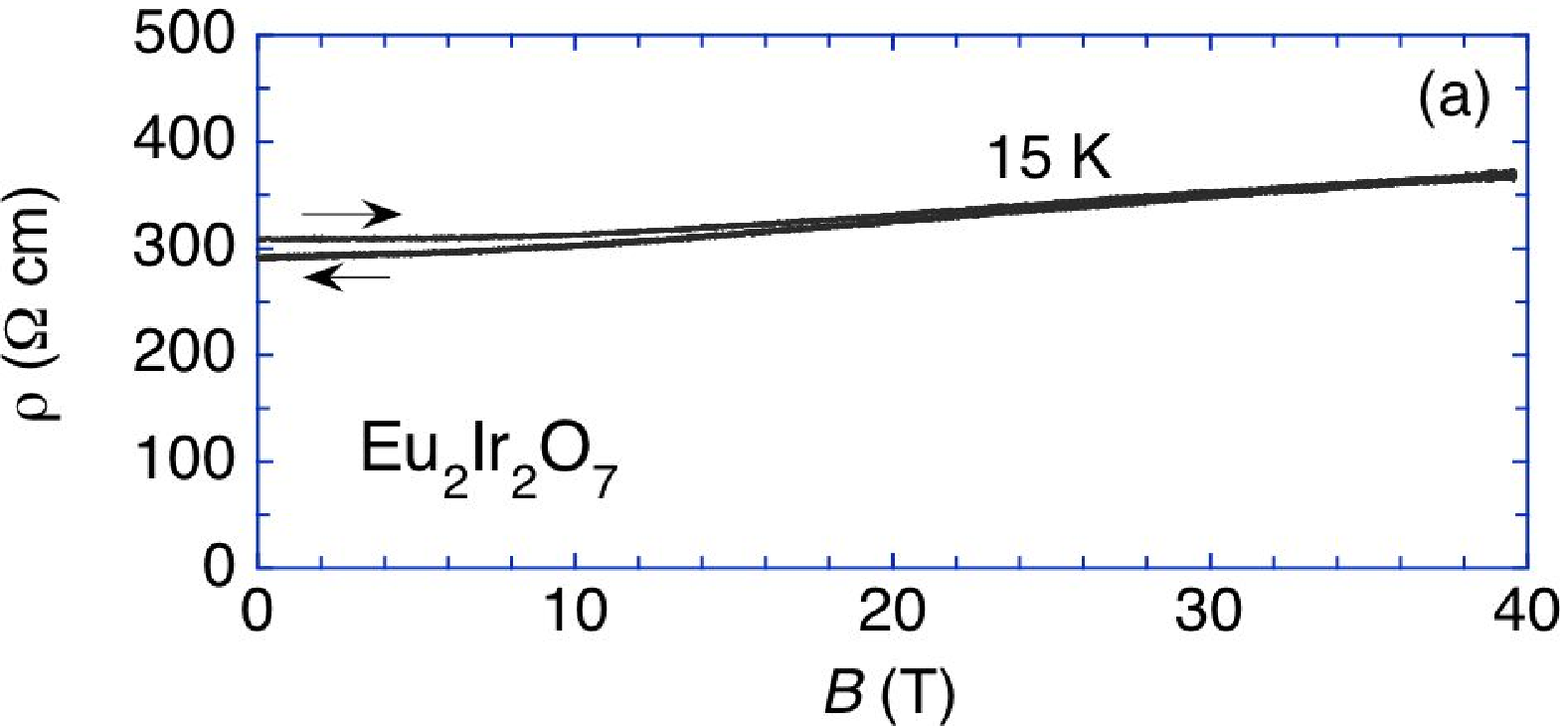}
\includegraphics[width=7.5cm]{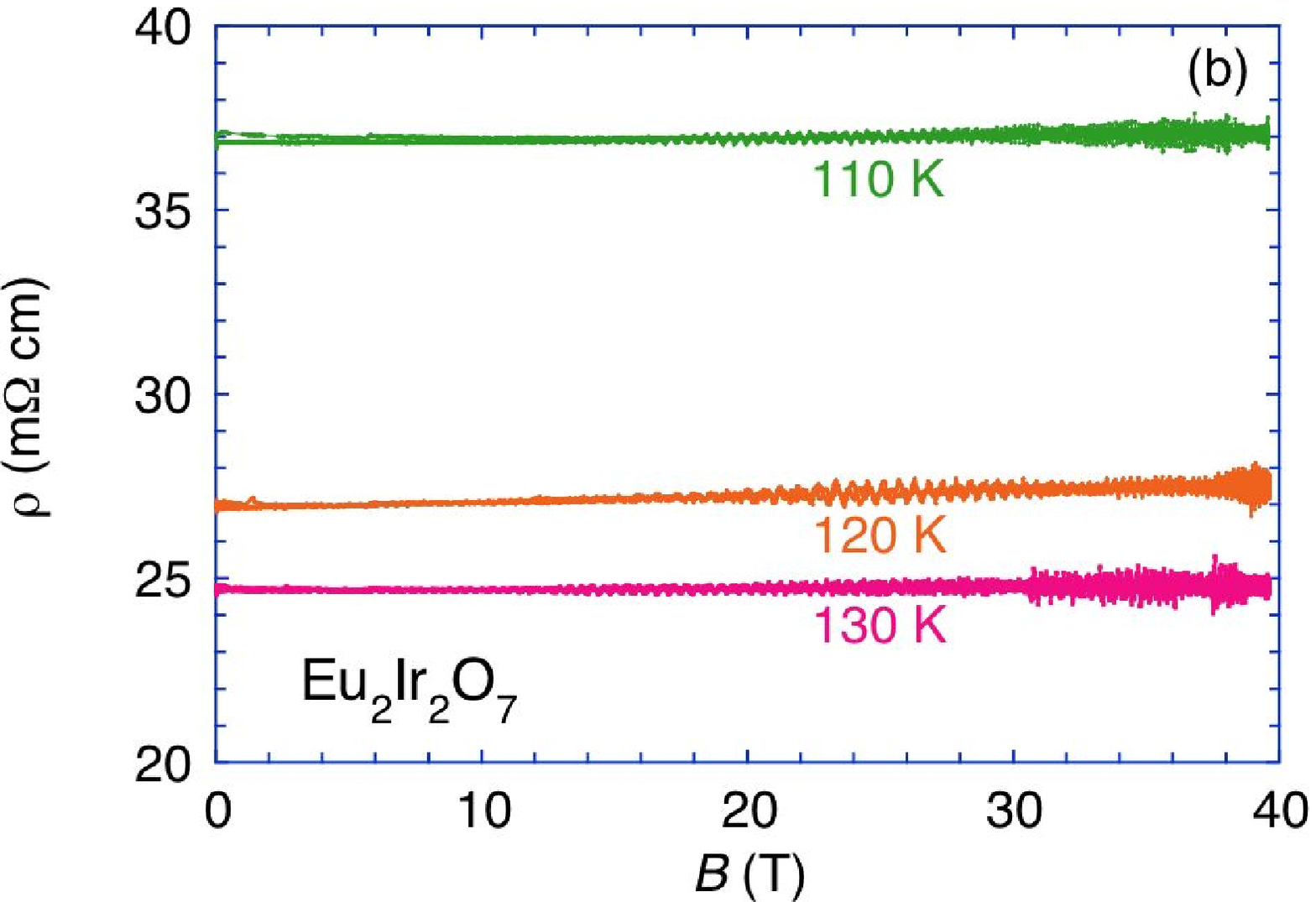}
\includegraphics[width=7.5cm]{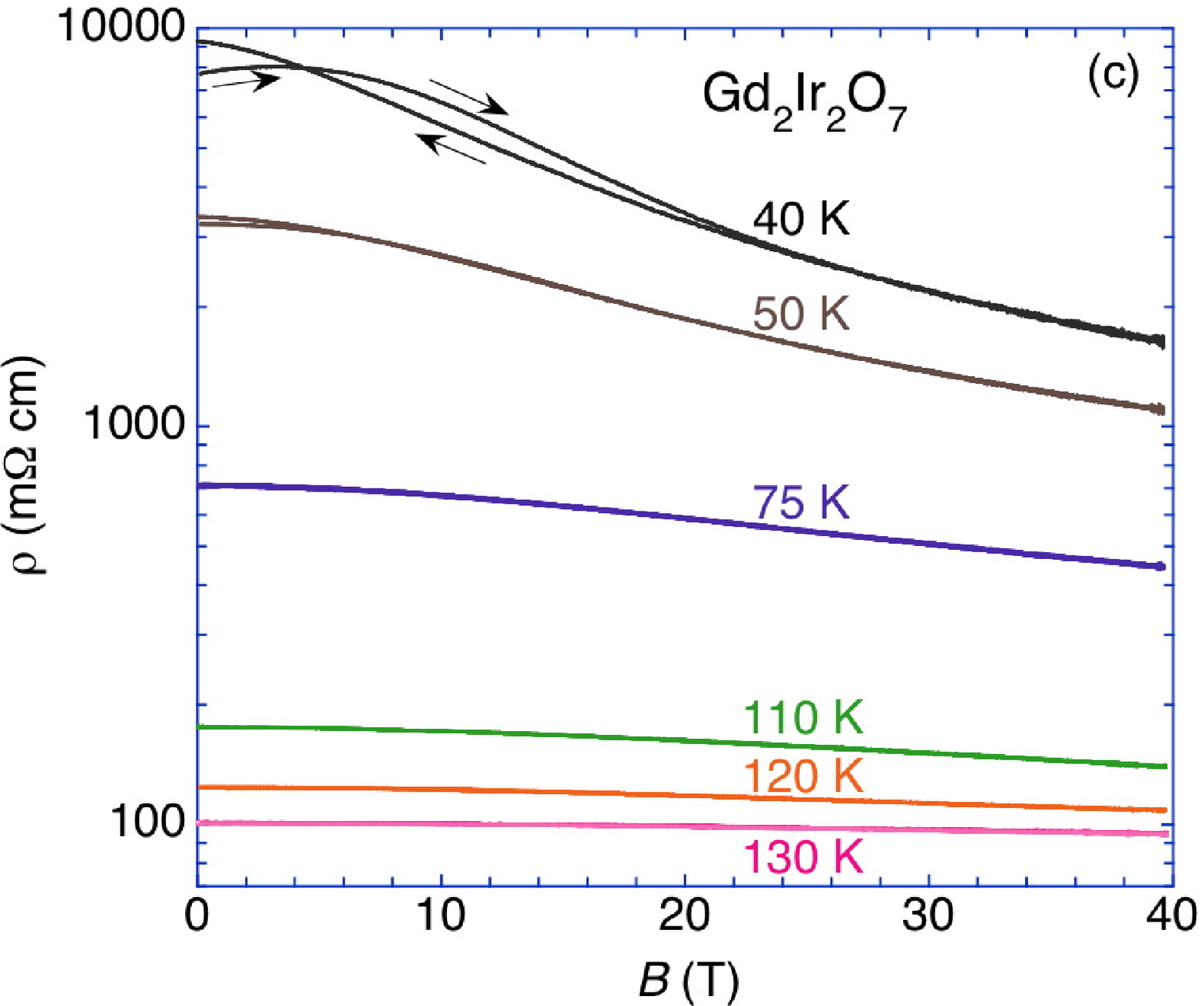}
\end{center}
\caption{
 (a) (Color online) Magnetoresistance of Eu$_2$Ir$_2$O$_7$ at 15 K under a pulsed magnetic field. (b) (Color online) Magnetoresistance of Eu$_2$Ir$_2$O$_7$ at 110, 120, and 130 K under a pulsed magnetic field. (c) (Color online) Magnetoresistance of Gd$_2$Ir$_2$O$_7$ under a pulsed magnetic field.
}
\label{f4}
\end{figure}

In summary, we investigated the MR effect of the pyrochlore oxide Nd$_2$Ir$_2$O$_7$, which exhibits an MIT at $T_{\rm MI}$ = 33 K.
In the metallic state above $T_{\rm MI}$, a small positive MR effect is observed, while a large negative MR effect is observed in the insulating state below $T_{\rm MI}$.
MR effects ($[\rho(B = 0) - \rho(B)]/\rho(B)$) exceeding 3000\% are found at 1 K at a field of 9 T.
We confirmed the crossover from the insulating state to a state with a small or partial band gap.
Furthermore, we revealed that the MR effect of the pyrochlore iridate {\it Ln}$_2$Ir$_2$O$_7$ depends on the magnetism of {\it Ln}$^{3+}$ ions.
The MR effect of Gd$_2$Ir$_2$O$_7$ ($T_{\rm MI}$ = 127 K) with magnetic Gd$^{3+}$ ions is large and negative.
On the other hand, the MR effect of Eu$_2$Ir$_2$O$_7$ ($T_{\rm MI}$ = 120 K) with nonmagnetic Eu$^{3+}$ ions is small and positive.
These results strongly indicate a significant role of $d$-$f$ interaction in the large negative MR effect.
A theoretical study would be desirable to explain the MR effect in {\it Ln}$_2$Ir$_2$O$_7$.

\section*{Acknowledgments}
We would like to thank S. Watanabe, M. Udagawa, and Y. Motome for their helpful discussions.
XRD measurements were performed at the Center for Instrumental Analysis at Kyushu Institute of Technology.
This work was supported by a Grant-in-Aid for Scientific Research on Priority Areas ''Novel States of Matter Induced by Frustration'' (No. 19052005) and Grants-in-Aid for Scientific Research on Innovation Areas ''Heavy Electrons'' (No. 21102518).
This research was partly supported by Grants-in-Aid for Scientific Research (C) (No. 23540417) and (B) (No. 23340096) from MEXT, Japan.


\begin{thebibliography}{99}
\bibitem{CC} The coupling constants are 1.0 and 2.2 for up- and down-sweeps, respectively. These values are smaller than those of the Mn perovskite.
\end{thebibliography}
\end{document}